# Why Do We See So Few Black Holes in Massive Binaries?

S. V. Karpov* and V. M. Lipunov

*Sternberg Astronomical Institute, Universitetskiĭ pr. 13, Moscow, 119899 Russia*
Received April 6, 2001

**Abstract**—We offer a simple explanation for the small number of black holes observed in pairs with massive stars. In detached massive binaries, spherically symmetric accretion takes place. This accretion could result in effective energy release in the hard band only if the equipartition of the gravitational and magnetic energy of plasma is established (Shvartsman's theorem). However, we show that due to the magnetic exhaust effect this equilibrium is virtually never established for the actual magnetic fields observed on massive stars: Shvartsman's theorem does not work. As a result, it is virtually impossible to detect black holes in detached massive binaries by currently available means (mainly, through X-ray observations). © *2001 MAIK "Nauka/Interperiodica"*.

Key words: *massive binaries, black holes*

## INTRODUCTION

Simple estimates based on the Salpeter function and on the assumption that black holes are formed from massive (~40–50$M_\odot$) stars predict the presence of ~50–100 million black holes in our Galaxy. Despite all the uncertainties and the development of new exotic scenarios for the formation of stellar-mass black holes, the actual number of black holes in the entire Galaxy cannot be fewer than $10^7$. Much of them must form pairs, for example, with massive stars. The expected number of massive binaries with black holes can be determined by multiplying the total number of black holes by the small ratio of the lifetime of a massive star to the Hubble time, i.e., $10^7 \times 10^6/10^{10} = 1000$. Here, we took the lifetime of a massive star, assuming that the optical companion (given the interchange of roles) is no less massive than the black-hole progenitor. Thus, one might expect ~1000 black-hole candidates in massive binaries. This number is in conflict with experimental data: we currently observe slightly more than ten candidates in binary systems. Remarkably, among these candidates only three belong to massive binaries, where they must be formed; moreover, some of them are not in our Galaxy. It should be added that the black-hole candidate Cygnus X-1 is a very close binary with an optical component that almost fills its Roche lobe. Clearly, there must be hundreds of times more detached binaries composed of a massive OB star and a stellar-wind accreting black hole. However, we do not see such systems.

Here, we draw attention to the fact that accretion in massive binary systems does not result in the effective generation of a hard radiation component, whose detection alone can point to the presence of a black-hole companion in the system. This is because the accretion in most systems of interest is spherically symmetric. This accretion can result in effective energy release in the hard band only if the equipartition is established between the magnetic and gravitational energies in the accretion flow.

However, the regime of magnetic exhaust operates in such systems (Lipunov 1987), which prevents the establishment of such equilibrium. As a result, most black holes in pairs with massive stars are unobservable.

## NO DISK IS FORMED IN DETACHED MASSIVE BINARIES WITH A BLACK HOLE

First, we note that an accretion disk is virtually never formed in detached massive binaries with stellar wind. The condition under which the matter captured by the black hole does not form a disk is the smallness of its mean momentum (see, e.g., Lipunov 1987)

$$\eta_t \Omega R_G^2 < \sqrt{3 r_g G M},$$
$$\Omega^2 = \frac{G(M + M_*)}{a^3},$$

where $\eta_t < 1/4$ and the radius of capture $R_G$ is determined from the equality of the wind kinetic and gravitational energies

$$\frac{1}{2}\rho V^2 = \frac{G M \rho}{R_G},$$
$$R_G = r_g \left(\frac{c}{V}\right)^2.$$

Accordingly, we constrain the velocity

---

* E-mail address for contacts: karpov@sai.msu.ru

$$V < V_{cr} = c\eta_t^{1/4}\left(\frac{1+q}{3q}\right)^{1/8} r_g^{3/8} a^{-3/8},$$

$$V_{cr} = 500\eta_{0.1}^{1/4}\left(\frac{1+q_1}{3q_1}\right)^{1/8} M_{10}^{3/8} a_{200}^{-3/8} \text{ km s}^{-1}.$$

Since the stellar-wind velocity far from massive stars is no less than 1500–2000 km s$^{-1}$, an accretion disk is clearly never formed in such systems.

Note that the situation in Cygnus X-1 is completely different. The optical component virtually fills its Roche lobe, and the wind has not yet gathered speed at the distance of the black-hole orbit (about two stellar radii); therefore, favorable conditions for the formation of an accretion disk arise.

Thus, we emphasize the first important distinction of detached massive pairs with a black hole: the accretion in them is spherical.

## WHAT CAN WE SEE IN BINARY SYSTEMS?

Because of the large optical luminosity of OB stars, the only evidence for the presence of a black-hole companion is hard X-ray radiation. The hard radiation from a black hole depends significantly on the accretion regime. Thus, for example, a classical accretion disk is optically thick and sinks slowly enough for the free–free radiation to become noticeable, which yields a considerable X-ray luminosity, while the situation for spherical accretion is much more complex.

In detached binaries, the rate of accretion from the stellar wind is always lower than the Eddington rate; i.e., the halo optical depth is very small. In the presence of even a very small initial field, the main energy release mechanism is electron synchrotron radiation. However, at typical accretion-flow temperatures, the synchrotron mechanism mainly produces optical radiation (Shvartsman 1971), which we cannot distinguish against the background of a normal star.

Which nonthermal mechanisms can produce a hard radiation component? Bremsstrahlung is ineffective (Shvartsman 1971); the inverse Compton effect would be important at an appreciable efficiency of energy release in the low-energy range and in the presence of a large number of hot electrons. However, for magnetic fields weaker than those for the equipartition, the efficiency of synchrotron energy release is low, and the hard Compton component is actually absent as well.

Note, however, that this situation radically differs from the advective accretion flow (ADAF) model, which has recently gained wide acceptance. Narayan's ADAF (Narayan and Yi 1995) is essentially a disk model; the efficiency of energy release in it is low even in the presence of a strong magnetic field because of the weak energy exchange between the radiating electrons and protons heated through viscous stresses.

In our case, however, the accreted matter has virtually no momentum, its motion is radial, the heating mechanism is compression (electrons and protons are heated equally), and, in general, the efficiency of energy release in the optical range does not need to be very low: only the hard radiation component is absent.

In the absence of equipartition, the conceivable generation mechanisms for the high-energy particles that could radiate in the hard band [turbulent acceleration, see Gruzinov and Quataert (1999); acceleration in fast reconnections, see Bisnovatyĭ-Kogan and Lovelace (2000)] do not work either.

Thus, we may say that the criterion for the absence of a hard radiation component and, accordingly, for a black hole being unobservable is whether an equipartition of the magnetic and gravitational energies has time to be established in the accretion flow.

## STELLAR-WIND MAGNETIC FIELD

The magnetic field frozen in the outflow of matter from the star (stellar wind) must have a quasi-radial structure (the tangential components are suppressed at the wind acceleration stage); therefore, we can take the following law of variations in magnetic-field strength

$$B(r) = B_*\left(\frac{r_*}{r}\right)^2,$$

where $B_*$ is the stellar surface magnetic field. Note that under this *a priori* assumption, the radial magnetic-field distribution does not depend (to within a small tangential component) on the stellar-wind acceleration law.

## NO EQUIPARTITION IS ESTABLISHED

We cannot detect a black hole with spherically symmetric accretion when there is not enough time for an equipartition between the magnetic and gravitational energies to be established in the accretion flow.

The ratio of the magnetic- and gravitational-energy densities in the accreted matter with a frozen-in field increases with decreasing distance from the black hole as

$$\kappa = \frac{B^2}{8\pi} \frac{r}{GM\rho} \propto r^{-3/2}.$$

Accordingly, in order for an equipartition not to be established down to radius $R_{eq} = r_{eq}/r_g$, it is necessary that within the stellar wind at the capture radius (where the wind kinetic energy is equal to the gravitational energy)

$$\kappa_0 < \left(\frac{R_G}{r_g}\right)^{-3/2} R_{eq}^{3/2} = \left(\frac{c}{V}\right)^{-3} R_{eq}^{3/2},$$

$$\kappa_0 \left(\frac{c}{V}\right)^3 < R_{eq}^{3/2},$$

$$\kappa_0 = \frac{B^2}{8\pi} \frac{2}{\rho V^2} = \frac{B^2}{8\pi} \frac{8\pi a^2}{\dot{M}_* V} = \frac{B^2 a^2}{\dot{M}_* c}\left(\frac{c}{V}\right) = \frac{B_*^2 r_*^4}{\dot{M}_* c a^2}\left(\frac{c}{V}\right).$$

Accordingly, the criterion for an equipartition not to be established in the accretion flow is

$$\frac{B_*^2 r_*^4}{\dot{M}_* c a^2}\left(\frac{c}{V}\right)^4 < R_{\rm eq}^{3/2}.$$

Normalizing the stellar radius to the solar radius, the outflow rate to $10^{-6} M_\odot$ year$^{-1}$, the binary's semimajor axis $200 R_\odot$, and the radius $r_{\rm eq}$ at which an equipartition is established to $3 r_{\rm g}$, we obtain

$$\left(\frac{B_*}{1\,{\rm G}}\right)^2 \left(\frac{r_*}{1\,R_\odot}\right)^4 \left(\frac{\dot{M}_*}{10^{-6} M_\odot/{\rm year}}\right)^{-1}$$
$$\times \left(\frac{a}{200 R_\odot}\right)^{-2} \left(\frac{1000\,{\rm km/s}}{V}\right)^4 \left(\frac{r_{\rm eq}}{3 r_{\rm g}}\right)^{-3/2} < 10^4. \quad (1)$$

Rewriting (1) as a condition on the initial magnetic field, we obtain

$$B_* < 1 \left(\frac{r_*}{10 R_\odot}\right)^{-2} \left(\frac{\dot{M}_*}{10^{-6} M_\odot/{\rm year}}\right)^{1/2} \left(\frac{a}{200 R_\odot}\right)$$
$$\times \left(\frac{1000\,{\rm km/s}}{V}\right)^{-2} \left(\frac{r_{\rm eq}}{3 r_{\rm g}}\right)^{3/4} {\rm G}. \quad (2)$$

We know almost nothing about the magnetic fields on OB stars. Until now, no strong magnetic field has been detected in any OB star (Mathys 1998). However, the low accuracy of such measurements (~200–300 G) does not allow us to draw conclusions about the presence of a weak field, which is of interest here. Nevertheless, some theoretical considerations (the absence of convection and, accordingly, of dynamo mechanisms for field enhancement) also suggest that the magnetism on these stars is weak. That may be why accreting black holes are unobservable in such systems.

## BLACK HOLES IN PAIRS WITH WOLF–RAYET STARS

As numerical simulations of the evolution of massive binaries show, a considerable fraction of binary systems with black holes survive after the second mass transfer, and pairs of helium stars (Wolf–Rayet stars) with black holes must be formed in nature. Their number must be an order of magnitude smaller than the number of such systems with OB stars, i.e., at least several tens in the Galaxy. Nevertheless, here, there is a conflict with observations as well. Apart from Cygnus X-3, no other black hole + Wolf–Rayet candidate systems are known at present.

Indeed, let us substitute the typical parameters of helium stars in (1) and (2). First, we take into account the fact that their radii are almost an order of magnitude smaller than those of OB stars, and, in general, their stellar-wind velocity (because of a higher escape velocity) is a factor of 2 or 3 higher, in some cases reaching 5000 km s$^{-1}$! This all makes the formation of an accretion disk in binary systems and the establishment of energy equipartition for spherical accretion unlikely, which may be responsible for the observed deficit of such systems.

## CONCLUSION

We have shown that for the typical parameters of a wide massive star–black hole binary system, spherical accretion takes place. Massive OB and WR stars do not have strong magnetic fields; because of magnetic exhaust, the field in the accretion flow is also weak (weaker than for equipartition), and the matter cannot produce hard radiation. Accordingly, we cannot see the black hole. We actually see no hard radiation sources in wide pairs with such stars.

## ACKNOWLEDGMENTS

We wish to thank V.S. Beskin, S.V. Bogovalov, and M.E. Prokhorov for fruitful discussions. This work was supported by the Russian Foundation for Basic Research (project no. 00-02-17164) and the State Science and Technology Program "Astronomy" (1.4.2.3).

*Translated by G. Rudnitskiĭ*